%% file: polyroot-rev.tex
\documentclass[leqno,draft,11pt]{article}
\usepackage{amssymb,amsmath,theorem,a4wide}
\usepackage[final]{graphicx}

\def\dedic{\thanks{Supported by grant IAA100190902 of GA AV \v CR,
project 1M0545 of M\v SMT \v CR, and RVO: 67985840.}}

\def\enumup{}
\input{defs}

\title{Root finding with threshold circuits}

\begin{document}
\maketitle
\begin{abstract}
We show that for any constant $d$, complex roots of degree $d$
univariate rational (or Gaussian rational) polynomials---given by a
list of coefficients in binary---can be computed to a given accuracy
by a uniform $\tc$ algorithm (a uniform family of constant-depth
polynomial-size threshold circuits). The basic idea is to compute the
inverse function of the polynomial by a power series. We also discuss
an application to the theory $\vtc$ of bounded arithmetic.
\end{abstract}
\section{Introduction}
The complexity class $\tc$ was originally defined by Hajnal et
al.~\cite{tc0} in the nonuniform setting, as the class of problems
recognizable by a family of polynomial-size constant-depth circuits
with majority gates. It was implicitly studied before by Parberry and
Schnitger~\cite{par-sch}, who consider various
models of computation using thresholds (threshold circuits, Boltzmann
machines, threshold RAM, threshold Turing machines). The importance of
the class follows already from the work of Chandra, Stockmayer, and
Vishkin~\cite{ac0red}, who show (in today's terminology) the
$\tc$-completeness of several basic problems (integer multiplication,
iterated addition, sorting) under $\Ac$ reductions. Barrington,
Immerman, and Straubing \cite{founif} establish that there is a robust
notion of fully uniform $\tc$. (We will use $\tc$ to denote this
uniform $\tc$, unless stated otherwise.)

We can regard $\tc$ as the natural complexity class of elementary
arithmetical operations: integer multiplication is $\tc$-complete, whereas
addition, subtraction, and ordering are in $\Ac\sset\tc$. The
exact complexity of division took some time to settle.
Wallace~\cite{wall:div} constructed division circuits of depth
$O((\log n)^2)$ 
and bounded fan-in (i.e., $\nc^2$). Reif~\cite{reif} improved this
bound to $O(\log n\,\log\log n)$. Beame, Cook, and Hoover~\cite{bch}
proved that division, iterated multiplication, and exponentiation
(with exponent given in unary) are $\tc$-reducible to each other, and
constructed $\ptime$-uniform $\tc$ circuits for these problems. Chiu,
Davida, and Litow~\cite{cdl} exhibited logspace-uniform $\tc$ circuits
for division, showing in particular that division is computable in
$\cxt L$. Finally, Hesse, Allender, and Barrington~\cite{hab} proved that
division (and iterated multiplication) is in uniform $\tc$.

Using these results, other related problems can be shown to be
computable in $\tc$, for example polynomial division, iterated
multiplication, and interpolation. In particular, using iterated
addition and multiplication of rationals, it is possible to
approximate in $\tc$ functions presented by sufficiently nice power
series, such as $\log$, $\exp$, $x^{1/k}$, and trigonometric
functions, see e.g.\ Reif~\cite{reif}, Reif and Tate~\cite{reif-tate},
Maciel and Th\'erien~\cite{mac-the:ser}, and Hesse et al.~\cite{hab}.

Numerical computation of roots of polynomials is one of the oldest
problems in mathematics, and countless algorithms have been devised to
solve it, both sequential and parallel. The most popular methods are
based on iterative techniques that successively derive closer and
closer approximations to a root (or, sometimes, to all the roots
simultaneously) starting from a suitable initial approximation. Apart
from the prototypical Newton--Raphson iteration, there are for
instance Laguerre's method~\cite[\S9.5]{num-rec}, Brent's
method~\cite[\S10.3]{num-rec}, the Durand--Kerner
method~\cite{durand,kerner}, the Jenkins--Traub
algorithm~\cite{jen-tra}, and many others. One can also reduce root
finding to matrix eigenvalue computation, for which there are
iterative methods such as the QR algorithm~\cite{gol-loan:mat}.
Another class of root-finding algorithms are divide-and-conquer
approaches: the basic idea is to recursively factorize the polynomial
by identifying a suitable contour (typically, a circle) splitting the
set of roots roughly in half, and recovering coefficients of the
factor whose roots fall inside the contour from the residue theorem by
numerical integration. Algorithms of this kind include
Pan~\cite{pan:85}, Ben-Or et al.~\cite{bfkt}, Neff~\cite{neff}, Neff
and Reif~\cite{neff-reif}, and Pan~\cite{pan:opt}, see
Pan~\cite{pan:hist} for an overview. These algorithms place root
finding in $\nc$: for example, the algorithm of \cite{pan:opt} can
find $n$-bit approximations to all roots of a polynomial of degree $d\le n$ in
time $O((\log n)^2(\log d)^3)$ using $O(nd^2(\log\log n)/(\log d)^2)$
processors on an EREW PRAM. (More specifically, Allender~\cite{all:ac}
mentions that root finding is known to be in the $\cxt{\#L}$
hierarchy, but not known to be in $\cxt{GapL}$.)

The purpose of this paper is to demonstrate that in the case of
constant-degree polynomials, we can push the complexity of root
finding down to uniform $\tc$ (i.e., constant time on polynomially
many processors on a TRAM, in terms of parallel complexity), as in the
case of elementary arithmetical operations. (This is clearly optimal:
already locating the unique root of a linear polynomial amounts to
division, which is $\tc$-hard.) As a corollary, the binary expansion
of any algebraic constant can be computed in uniform $\tc$ when given
the bit position in unary. Our primary interest is theoretical, we
seek to investigate the power of the complexity class $\tc$; we do not
expect our algorithm to be competitive with established methods in
practice, and we did not make any effort to optimize parameters of the
algorithm.

The basic idea of the algorithm is to express the inverse
function of the polynomial by a power series, whose partial sums can
be computed in $\tc$ using the results of Hesse et al.~\cite{hab}. We
need to ensure that coefficients of the series are $\tc$-computable,
we need bounds on the radius of convergence and convergence rate of
the series, and we need to find a point in whose image to put the
centre of the series so that the disk of convergence includes the
origin. Doing the latter directly is in fact not much easier than
approximating the root in the first place, so we instead construct a
suitable polynomial-size set of sample points, and we invert the
polynomial at each one of them in parallel.

We formulated our main result in terms of computational complexity,
but our original motivation comes from logic (proof complexity). The
bounded arithmetical theory $\vtc$ (see Cook and
Nguyen~\cite{cook-ngu}), whose provably total computable functions are
the $\tc$ functions, can define addition, multiplication, and ordering
on binary integers, and it proves that these operations obey the basic
identities making it a discretely ordered ring. The question is which
other properties of the basic arithmetical operations are provable in
the theory, and in particular, whether it can prove induction (on
binary integers) for some class of formulas. Now, it follows easily
from known algebraic characterizations of induction for open formulas
in the language of ordered rings ($\io$, see Shepherdson~\cite{sheph})
and from the witnessing theorem for $\vtc$ that $\vtc$ proves $\io$ if
and only if for each $d$ there is a $\tc$ root-finding algorithm for
degree $d$ polynomials whose soundness is provable in $\vtc$. Our
result thus establishes the computational prerequisites for proving
open induction in $\vtc$, leaving aside the problem of formalizing the
algorithm in the theory. Since the soundness of the algorithm can be
expressed as a universal sentence, we can also reformulate this result
as follows: the theory $\vtc+\Th_{\forall\Sig0}(\stm)$ proves $\io$.

The paper is organized as follows. In Section~\ref{sec:preliminaries}
we provide some background in the relevant parts of complexity theory
and complex analysis. Section~\ref{sec:invert-poly} contains
material on inverting polynomials with power series.
Section~\ref{sec:root-finding-tc0} presents our main result, a $\tc$
root-finding algorithm.
Finally, in Section~\ref{sec:open-induction} we discuss the connection
to bounded arithmetic.

\section{Preliminaries}\label{sec:preliminaries}
A language $L$ is in \emph{nonuniform $\tc$} if there is a sequence of
circuits $C_n\colon\{0,1\}^n\to\{0,1\}$ consisting of unbounded fan-in
majority and negation gates such that $C_n$ computes the
characteristic function of $L$ on strings of length $n$, and $C_n$ has
size at most $n^c$ and depth $c$ for some constant $c$.

$L$ is in \emph{(uniform) $\tc$,} if the sequence
$\{C_n:n\in\omega\}$ is additionally $\cxt{DLOGTIME}$-uniform
($U_D$-uniform in the terminology of Ruzzo~\cite{ruzzo}): i.e., we can
enumerate the gates in the circuit by numbers $i<n^{O(1)}$ in such a
way that one can check the type of gate $i$ and whether gate $i$ is an
input of gate $j$ by a deterministic Turing machine in time $O(\log
n)$, given $n,i,j$ in binary. There are other equivalent
characterizations of $\tc$. For one, it coincides with languages
recognizable by a threshold Turing machine \cite{par-sch} in
time $O(\log n)$ with $O(1)$ thresholds \cite{all:tc0}.
Another important characterization is in terms of descriptive
complexity. We can represent a string $x\in\{0,1\}^n$ by the
first-order structure $\cmb\p{\{0,\dots,n-1\},{<},\bit,X}$, where $X$
is a unary predicate encoding the bits of $x$. Then a language is in
$\tc$ iff its corresponding class of structures is definable by a
sentence of $\cxt{FOM}$ (first-order logic with majority quantifiers). We
refer the reader to \cite{founif} for more background on uniformity of
$\tc$.

In some cases it may be more convenient to consider languages in a
non-binary alphabet $\Sigma$. The definition of $\tc$ can be adapted
by adjusting the input alphabet of a threshold Turing machine, or by
considering more predicates in the descriptive complexity
setting. In the original definition using threshold circuits, the same
can be accomplished by encoding each symbol of $\Sigma$ with a binary
substring of fixed length. We can also define $\tc$ predicates with more
than one input in the obvious way.

A function $f\colon\{0,1\}^*\to\{0,1\}^*$ is computable in $\tc$ if
the length of its output is polynomially bounded in the length of its
input, and its bitgraph is a $\tc$ predicate. (The bitgraph of $f$ is
a binary predicate $b(x,i)$ which holds iff the $i$th bit of $f(x)$ is
$1$.) In terms of the original definition, this amounts to allowing
circuits $C_n\colon\{0,1\}^n\to\{0,1\}^{m(n)}$, where $m(n)=n^{O(1)}$.
$\tc$ functions are closed under composition, and under ``parallel
execution'': if $f$ is a $\tc$ function, its aggregate function
$g(\p{x_0,\dots,x_{m-1}})=\p{f(x_0),\dots,f(x_{m-1})}$ is also in~$\tc$.
We note in this regard that $\tc$ functions can do basic
processing of lists
$$x_0,x_1,\dots,x_{m-1}$$
where ``,'' is a separator character. Using the fact that $\tc$ can
count commas (and other symbols), we can for instance extract the
$i$th element from the list, convert the list to and from a
representation where each element is padded to some fixed length with
blanks, or sort the list according to a given $\tc$ comparison
predicate.

We will refrain from presenting $\tc$ functions in one of the
formalisms suggested by the definitions above: we will give informal
algorithms, generally consisting of a constant number of simple steps
or $\tc$ building blocks, sometimes forking into polynomially many
parallel threads. The reader should have no difficulty convincing
herself that our algorithms are indeed in~$\tc$.

We will work with numbers of various kinds. Integers will be
represented in binary as usual, unless stated otherwise. As we already
mentioned in the introduction, elementary arithmetical operations on
integers are $\tc$ functions: this includes addition, subtraction,
ordering, multiplication, division with remainder, exponentiation
(with unary exponents), iterated addition, iterated multiplication,
and square root approximation. Here, iterated addition is the function
$\p{x_0,\dots,x_{m-1}}\mapsto\sum_{i<m}x_i$, and similarly for
multiplication. Notice that using iterated multiplication, we can also
compute factorials and binomial or multinomial coefficients of unary
arguments. Base conversion is also in~$\tc$.

Rational numbers will be represented as pairs of integers, indicating
fractions. We cannot assume fractions to be reduced, since integer
$\gcd$ is not known to be $\tc$-computable. Using integer division, we
can convert a fraction to its binary expansion with a given accuracy
(the opposite conversion is trivial). Rational arithmetic is reducible
to integer arithmetic in the obvious way, hence rational addition,
subtraction, ordering, multiplication, division, exponentiation (with
unary integer exponents), iterated addition, iterated
multiplication, and square root approximation are in~$\tc$.

In lieu of complex numbers, we will compute with Gaussian rationals
(elements of the field $\Q(i)$), represented as pairs of rationals
$a+ib$. By reduction to rational arithmetic, we can see that addition,
subtraction, complex conjugation, norm square, norm approximation,
multiplication, division, and iterated addition of Gaussian rationals
are in~$\tc$. Using the binomial theorem, exponentiation with unary
integer exponents is also in~$\tc$. (In fact, iterated multiplication of
Gaussian rationals is in~$\tc$ using conversion to polar coordinates, but
we will not need this.)

We will need some tools from complex analysis. We refer the reader to
Ahlfors~\cite{ahlf} or Conway~\cite{conw:i} for background, however,
we review here some basic facts to fix the notation. A function
$f\colon U\to\C$, where $U\sset\C$ is open, is \emph{holomorphic} (or
\emph{analytic}) in $U$ if $f'(a)=\lim_{z\to a}(f(z)-f(a))/(z-a)$
exists for every $a\in U$. The set of all functions holomorphic in $U$
is denoted $H(U)$. Let $B(a,r):=\{z:\lh{z-a}<r\}$ and $\ob
B(a,r):=\{z:\lh{z-a}\le r\}$. If $f$ is holomorphic in the open disk
$B(a,R)$, it can be expressed by a \emph{power series}
$$f(z)=\sum_{n=0}^\infty c_n(z-a)^n$$
on $B(a,R)$. More generally, if $f$ is holomorphic in the annulus
$A=B(a,R)\bez\ob B(a,r)$, $0\le r<R\le\infty$, it can be written in
$A$ as a \emph{Laurent series}
$$f(z)=\sum_{n=-\infty}^{+\infty}c_n(z-a)^n.$$
We denote the coefficients of the series by $[(z-a)^n]f:=c_n$. (Other
variables may be used instead of $z$ when convenient.) The
\emph{residue} of $f$ at $a$ is $\Res(f,a):=[(z-a)^{-1}]f$. When
$a=0$, we write just $[z^n]f$ and $\Res(f)$, respectively. The coefficients of a
Laurent series are given by \emph{Cauchy's integral formula}:
$$[(z-a)^n]f=\frac1{2\pi i}\int_\gamma\frac{f(z)}{(z-a)^{n+1}}\,dz,$$
where $\gamma$ is any closed curve in $A$ whose index with
respect to $a$ is $1$ (such as the circle $\gamma(t)=a+\roo e^{2\pi
it}$, $r<\roo<R$).
The \emph{identity theorem} states that if $f,g$ are holomorphic
in a region (i.e., connected open set) $U$ and coincide on a set
$X\sset U$ which has a limit point in $U$, then $f=g$.
The \emph{open mapping theorem} states that a
nonconstant function $f$ holomorphic in a region is an open mapping (i.e., maps open sets to open sets).

If $X\sset\C$ and
$a\in\C$, we put $\dist(a,X)=\inf\{\lh{z-a}:z\in X\}$.

We will also need some easy facts on zeros of polynomials. Let
$f\in\C[x]$ be a degree $d$ polynomial, and write
$f(x)=\sum_{j=0}^da_jx^j$. \emph{Cauchy's bound} \cite[L.~6.2.7]{cky} states that every
zero $\alpha$ of $f$ satisfies
$$\frac{\lh{a_0}}{\lh{a_0}+\max\limits_{0<j\le d}\lh{a_j}}\le\lh\alpha
\le1+\max_{j<d}\frac{\lh{a_j}}{\lh{a_d}}.$$
Let $f,g\in(\Q(i))[x]$ be two polynomials of degrees $d,e$ (resp.),
and assume $f(\alpha)=g(\beta)=0$, $\alpha\ne\beta$. If
$f,g\in(\Z[i])[x]$, we have
\[\tag{$*$}\lh{\alpha-\beta}\ge\frac1{(2^{d+1}\|f\|_\infty)^e\|g\|_2^d},\]
where $\|f\|_p$ denotes the $L_p$-norm of the vector of coefficients
of $f$ \cite[\S6.8]{cky}. In general, we can apply $(*)$ to the
polynomials $rf$ and $sg$, where $r$ is the product of all
denominators appearing among the coefficients of $f$, and similarly
for $s$. If we represent $f$ and $g$ by the lists of their
coefficients, which are in turn represented by quadruples of binary
integers as detailed above, we obtain easily the following root
separation bound:
\begin{Lem}\th\label{lem:sep}
For each $j=0,1$, let $f_j\in(\Q(i))[x]$ have degree $d_j$ and total
bit size $n_j$, and assume $f_j(\alpha_j)=0$. If
$\alpha_0\ne\alpha_1$, then
$$\lh{\alpha_0-\alpha_1}\ge2^{-(d_1n_0+d_0n_1)}\ge2^{-n_0n_1}.$$
\end{Lem}

\section{Inverting polynomials}\label{sec:invert-poly}
As already mentioned in the introduction, the main strategy of our
algorithm will be to approximate a power series computing the inverse
function of the given polynomial $f$. In this section, we establish
the properties of such series needed to make the algorithm work.

The basic fact we rely on is that holomorphic functions with
nonvanishing derivative are locally invertible: i.e., if $f\in H(U)$
and $a\in U$ is such that $f'(a)\ne0$, there exist open neighbourhoods
$a\in U_0\sset U$ and $f(a)\in V_0$ such that $f$ is a homeomorphism
of $U_0$ onto $V_0$, and the inverse function $g=(f\res U_0)^{-1}$ is
holomorphic in $V_0$. In particular, $g$ is computable by a power
series in a neighbourhood of $f(a)$.

Notice that local inverses of holomorphic functions are automatically
two-sided: if $f\in H(U)$, $g\in H(V)$, $a\in U$, $b\in V$,
$g(b)=a$, and $f(g(z))=z$ in a neighbourhood of $b$, then $g(f(z))=z$
in a neighbourhood of $a$.

The coefficients of the power series of an inverse of a holomorphic
function are given by the \emph{Lagrange inversion formula} {\cite[\S3.8,
Thm.~A]{comtet}}:
\begin{Fact}\th\label{fact:lagr}
Let $f\in H(U)$, $g\in H(V)$, $f\circ g=\id_V$, $a=g(b)\in U$,
$b=f(a)\in V$, $n>0$. Then
$$[(w-b)^n]g(w)=\frac1n\Res\left(\frac1{(f(z)-b)^n},a\right).$$
\end{Fact}
We can make the formula even more explicit as follows. First, the
composition of two power series is given by \emph{Fa\`a di Bruno's
formula} \cite[\S3.4, Thm.~A]{comtet}, which we formulate only for
$a=b=0$ for simplicity:
\begin{Fact}\th\label{fact:faa}
Let $f\in H(U)$, $g\in H(V)$, $g(0)=0\in U$, $f(0)=0\in V$, $n\ge0$. Then
$$[z^n](g\circ f)=
\sum_{\sum_{j=1}^\infty jm_j=n}
  \binom{\sum_jm_j}{m_1,m_2,\dots}[w^{\sum_jm_j}]g
  \prod_{j=1}^\infty\left([z^j]f\right)^{m_j}.$$
\end{Fact}
Note that here and below, the outer sum is finite, and the product has
only finitely many terms different from $1$, hence the right-hand side
is well-defined without extra assumptions on convergence.
We can now expand the residue in \th\ref{fact:lagr} to obtain the
following version of Lagrange inversion formula, which only refers to
the coefficients of $f$ \cite[\S3.8, Thm.~E]{comtet}:
\begin{Prop}\th\label{thm:invser}
Let $f\in H(U)$, $g\in H(V)$, $f\circ g=\id_V$, $a=g(b)\in U$,
$b=f(a)\in V$. Then $[(w-b)^0]g=a$, and for $n>0$,
$$[(w-b)^n]g=
  \frac1{n!\,[z-a]f}
  \sum_{\sum_{j=2}^\infty(j-1)m_j=n-1}
   \left(\textstyle\sum_jjm_j\right)!\,
   \prod_{j=2}^\infty\frac1{m_j!}
         \left(-\frac{[(z-a)^j]f}{([z-a]f)^j}\right)^{m_j}.
$$
\end{Prop}
\begin{Pf}
Note that $f'(a)\ne0$. Put $f_1(z)=f(a+z/f'(a))-b$ and
$g_1(w)=f'(a)(g(b+w)-a)$, so that $f_1(0)=0=g_1(0)$, $f_1\circ g_1=\id$
on a neighbourhood of~$0$, and $f'_1(0)=1$. 
Write $f_1(z)=z(1-h(z))$, where $h$ is holomorphic in a neighbourhood
of~$0$, and $h(0)=0$.
Then
\begin{align*}
[w^n]g_1=\frac1n[z^{-1}]\frac1{f_1^n}&=
\frac1n[z^{n-1}]\frac1{(1-h)^n}\\
&=\frac1n\sum_{\sum_{j=1}^\infty jm_j=n-1}
 \frac{(\sum_jm_j)!}{m_1!\,m_2!\,\cdots}
 \frac{(\sum_jm_j+n-1)!}{(\sum_jm_j)!\,(n-1)!}
 \prod_{j=1}^\infty\left([z^j]h\right)^{m_j}\\
&=\frac1{n!}\sum_{\sum_{j=1}^\infty jm_j=n-1}
 \frac{(\sum_j(j+1)m_j)!}{m_1!\,m_2!\,\cdots}
 \prod_{j=1}^\infty\left(-[z^{j+1}]f_1\right)^{m_j}\\
&=\frac1{n!}\sum_{\sum_{k=2}^\infty(k-1)m_k=n-1}
 \left(\textstyle\sum_kkm_k\right)!\,
 \prod_{k=2}^\infty\frac{\left(-[z^k]f_1\right)^{m_k}}{m_k!}
\end{align*}
using Facts \ref{fact:lagr} and~\ref{fact:faa}, and the expansion
$[w^r](1-w)^{-n}=\binom{r+n-1}{n-1}$. The result follows by noting
that for any $k,n>0$, $[z^k]f_1=([(z-a)^k]f)/(f'(a))^k$, $[w^n]g_1=f'(a)[(w-b)^n]g$,
and $f'(a)=[z-a]f$.
\end{Pf}

Let $d$ be a constant. If $f$ in \th\ref{thm:invser} is a polynomial of degree $d$,
then the product is nonzero only when $m_j=0$ for every $j>d$, hence
it suffices to enumerate $m_2,\dots,m_d$. It follows easily that
the outer sum has polynomially many (namely, $O(n^d)$) terms, and we
can compute $[(w-b)^n]g$ in uniform $\tc$ given $a$, $b$, and the
coefficients of $f$ in binary, and $n$ in unary.

Apart from a description of the coefficients, we also need bounds on
the radius of convergence of the inverse series, and on its rate of
convergence (i.e., on the norm of its coefficients). Generally
speaking, the radius of convergence of a power series is the distance
to the nearest singularity. Since a polynomial $f$ is an entire proper
map, its inverse cannot escape to infinity or hit a point where $f$ is
undefined, thus the only singularities that can happen are
branch points. These occur at zeros of $f'$. This suggests that the
main parameter governing the radius of convergence and other
properties of the inverse should be the distance of $a$ to the set
$C_f=\{z\in\C:f'(z)=0\}$ of \emph{critical points} of $f$.
\begin{Lem}\th\label{lem:inj}
Let $f\in\C[x]$ be a degree $d$ polynomial with no roots in $B(a,R)$,
$R>0$, and let $\mu>0$.
Then $\lh{f(z)-f(a)}<\bigl((1+\mu)^d-1\bigr)\lh{f(a)}$ for all $z\in B(a,\mu R)$.
\end{Lem}
\begin{Pf}
Write $f(z)=c\prod_{j=1}^d(z-\alpha_j)$. We have
$$\frac{f(z)}{f(a)}=\prod_{j=1}^d\frac{(z-a)+(a-\alpha_j)}{a-\alpha_j}
 =\sum_{I\sset\{1,\dots,d\}}\prod_{j\in I}\frac{z-a}{a-\alpha_j},$$
hence
$$\left|\frac{f(z)}{f(a)}-1\right|
=\Bigl|\sum_{I\ne\nul}\prod_{j\in I}\frac{z-a}{a-\alpha_j}\Bigr|
\le\sum_{I\ne\nul}\prod_{j\in I}\frac{\lh{z-a}}R
<(1+\mu)^d-1.\qedhere$$
\end{Pf}
\begin{Prop}\th\label{thm:radius}
Let $f\in\C[x]$ have degree $d>1$, $f(a)=b$,
and $0<R\le\dist(a,C_f)$. Let
$$g(w)=a+\sum_{n=1}^\infty c_n(w-b)^n$$
satisfy $f\circ g=\id_{B(b,\roo)}$, where $\roo>0$ is the radius of
convergence of $g$. Put
\begin{align*}
\mu&=\sqrt[d-1]2-1\ge\frac{\ln2}{d-1},&
\nu&=\frac{2(d-1)\mu-1}d\ge\frac{\ln 4-1}d,\\
\lambda_\delta&=\sqrt[d\,]{1+\delta d\nu}-1\ge\frac{\delta\ln\ln 4}d,&
\roo_0&=\nu R\,\lh{f'(a)}
\end{align*}
for $0<\delta\le1$ (the inequalities are established below). Then:
\begin{enumerate}
\item\label{item:inj} $f$ is injective on $\ob B(a,\mu R)$.
\item\label{item:conv} $\roo\ge\roo_0$.
\item\label{item:lift} $g[B(b,\roo)]\Sset B(a,\lambda_1R)$, and more
generally, $g[B(b,\delta\roo_0)]\Sset B(a,\lambda_\delta R)$ for each $\delta\in(0,1]$.
\item\label{item:coef} $\lh{c_n}\le\mu R/n\roo_0^n$.
\end{enumerate}
\end{Prop}
\begin{Pf}
Notice that $e^x-1\ge x$ for every $x\in\RR$,
hence $\sqrt[d-1]2-1=\exp((\ln 2)/(d-1))-1\ge(\ln 2)/(d-1)$;
$\nu\ge(\ln4-1)/d$ immediately follows.
Similarly, $\lambda_\delta\ge\ln(1+\delta(\ln4-1))/d$. We have
$\ln(1+\delta(\ln4-1))\ge\delta\ln\ln4$ for $\delta\in[0,1]$ as $\ln$
is concave.

\eqref{item:inj}: Let $u,v\in\ob B(a,\mu R)$, $u\ne v$. We have
$$f(v)-f(u)=\int_u^vf'(z)\,dz
=(v-u)\left(f'(a)+\int_0^1f'\bigl((1-t)u+tv\bigr)-f'(a)\,dt\right).$$
Since $\lh{f'((1-t)u+tv)-f'(a)}<\lh{f'(a)}$ for all $t\in(0,1)$ by
\th\ref{lem:inj}, we obtain
$$\left|\int_0^1f'\bigl((1-t)u+tv\bigr)-f'(a)\,dt\right|
\le\int_0^1\left|f'\bigl((1-t)u+tv\bigr)-f'(a)\right|\,dt
<\lh{f'(a)},$$
thus $f(u)\ne f(v)$.

\eqref{item:conv}: Let $U=B(a,\mu R)$. Since $f$ is a biholomorphism
of $U$ and $f[U]$, $\roo\ge\dist(b,\C\bez f[U])$. Since $f[U]$ is
open, there exists $w\notin f[U]$ such that $\lh{w-b}=\dist(b,\C\bez
f[U])$. Let $z_n\in U$ be such that $\lim_nf(z_n)=w$. By compactness,
$\{z_n\}$ has a convergent subsequence; without loss of generality,
there exists $z=\lim_nz_n$. Then $f(z)=w$ by continuity, hence
$z\notin U$. However, $z\in\ob U$, hence $z$ is in the topological
boundary $\partial U=\ob U\bez\Int U=\ob U\bez U$. We have
thus verified that $\roo\ge\dist(b,f[\partial U])$.

Let $u=a+\mu Re^{i\theta}\in\partial U$. We have
$$f(u)=b+\int_a^uf'(z)\,dz
=b+Re^{i\theta}\left(\mu f'(a)+\int_0^\mu f'(a+te^{i\theta}R)-f'(a)\,dt\right).$$
By \th\ref{lem:inj},
$\lh{f'(a+te^{i\theta}R)-f'(a)}\le\bigl((1+t)^{d-1}-1\bigr)\lh{f'(a)}$, hence
\begin{multline*}
\left|\int_0^\mu f'(a+te^{i\theta}R)-f'(a)\,dt\right|
\le\lh{f'(a)}\int_0^\mu(1+t)^{d-1}-1\,dt
=\lh{f'(a)}\left(\frac{(1+\mu)^d-1}d-\mu\right)\\
=\lh{f'(a)}\,\frac{2(1+\mu)-1-d\mu}d
=\lh{f'(a)}\,\frac{1-(d-2)\mu}d.
\end{multline*}
Thus,
$$\lh{f(u)-b}\ge R\,\lh{f'(a)}\left(\mu-\frac{1-(d-2)\mu}d\right)
=\nu R\,\lh{f'(a)}.$$

\eqref{item:lift}: The proof above shows that $g[B(b,\roo_0)]\sset U$.
As $f$ is injective on $U\Sset B(a,\lambda_\delta R)$, it suffices to
show that $f[B(a,\lambda_\delta R)]\sset B(b,\delta\roo_0)$. Let
thus $u=a+\lambda Re^{i\theta}$, $\lambda<\lambda_\delta$. As above,
$$f(u)=b+Re^{i\theta}
  \left(\lambda f'(a)+\int_0^\lambda f'(a+te^{i\theta}R)-f'(a)\,dt\right)$$
and
$$
\left|\int_0^\lambda f'(a+te^{i\theta}R)-f'(a)\,dt\right|
\le\lh{f'(a)}\left(\frac{(1+\lambda)^d-1}d-\lambda\right),$$
hence
$$
\lh{f(u)-b}
\le R\,\lh{f'(a)}\frac{(1+\lambda)^d-1}d
<R\,\lh{f'(a)}\frac{(1+\lambda_\delta)^d-1}d=\delta\roo_0.
$$

\eqref{item:coef}: Let $\gamma(t)=a+\mu Re^{2\pi it}$. By
\th\ref{fact:lagr} and Cauchy's integral formula,
$$c_n=\frac1{2\pi in}\int_\gamma\frac{dz}{(f(z)-b)^n}
=\frac{\mu R}n\int_0^1\frac{e^{2\pi it}\,dt}{(f(\gamma(t))-b)^n}.$$
The proof of \eqref{item:conv} shows $\lh{f(\gamma(t))-b}\ge\roo_0$,
hence
$$\lh{c_n}\le\frac{\mu R}n\int_0^1\frac{dt}{\lh{f(\gamma(t))-b}^n}
\le\frac{\mu R}{n\roo_0^n}.\qedhere$$
\end{Pf}
\begin{Exm}
Let $f(z)=z^d$, $a=b=1$. Then $f'=dz^{d-1}$,
$C_f=\{0\}$, $R=1$, $f'(a)=d$. It is not hard to see that $f$ is
injective on $B(1,r)$ iff no two points of $B(1,r)$ have arguments
differing by $2\pi/d$ iff $r\le\sin(\pi/d)=\pi/d+O(d^{-3})$.
Since $g$ must hit a root of $f'$ at
the circle of convergence, we must have $\roo=1=(1/d)Rf'(a)$. Finally,
$\lh{(1+z)^d-1}$ is maximized on $\{z:\lh z=r\}$ for $z$ positive
real, thus $B(1,\lambda R)\sset g[B(1,\delta\roo)]$ iff $(1+\lambda)^d-1\le\delta$
iff $\lambda\le(1+\delta)^{1/d}-1=\ln(1+\delta)/d+O(d^{-2})$. Thus, in
\th\ref{thm:radius}, $\mu$, $\nu$, and $\lambda_\delta$ are optimal up
to a linear factor.
\end{Exm}
\begin{Rem}\th\label{rem:refs}
We prefer to give a simple direct proof of \th\ref{thm:radius} for the
benefit of the reader. Nevertheless, we could have assembled the
bounds (with somewhat different constants) from several more
sophisticated results in the literature. The Grace--Heawood theorem
(or rather its corollary, originally due to Alexander, Kakeya, and
Szeg\H o; see \cite[Thm.~23,2]{mard:poly}) states that
\eqref{item:inj} holds with $\mu=\sin(\pi/d)$ (which is tight in view
of the $z^d$ example). Then the Koebe $1/4$-theorem
\cite[Thm.~14.7.8]{conw:ii} implies \eqref{item:conv} with
$\nu=\mu/4$, and one more application of the theorem yields
\eqref{item:lift} with $\lambda_\delta=\nu\delta/4$.
\end{Rem}
\section{Root finding in $\tc$}\label{sec:root-finding-tc0}
We start with the core part of our root-finding algorithm. While it is
conceptually simple, its output is rather crude, so we will have to
combine it with some pre- and postprocessing to obtain the desired
result (\th\ref{thm:root}).
\begin{Thm}\th\label{thm:list}
Let $d$ be a constant. There exists a uniform $\tc$ function which,
given the coefficients of a degree $d$ polynomial $f\in(\Q(i))[x]$ in
binary and $t$ in unary, computes a list
$\{z_j:j<s\}\sset\Q(i)$ such that every complex root of $f$ is within
distance $2^{-t}$ of some $z_j$.
\end{Thm}
\begin{Pf}
If $d=1$, it suffices to divide the coefficients of~$f$. Assume $d\ge2$. Let
$\mu,\nu,\lambda=\lambda_{1/2}$ be as in \th\ref{thm:radius} (more
precisely, we should use their fixed rational approximations; we will
ignore this for simplicity). Let $A=1+\lambda/5$,
$p=\cl{5\pi/\lambda}$, and $\xi=e^{2\pi i/p}$ (approximately, again).
Consider the $\tc$ algorithm given by the following
description:
\begin{enumerate}
\item Input: $f=\sum_{j\le d}f_jz^j$ with $f_j\in\Q(i)$, $f_d\ne0$,
and $t>0$ in unary.
\item\label{item:rec} Put $\ep=2^{-t}$. Compute recursively a list
$C=\{\alpha_j:j<s\}$ including $\ep/4$-approximations of all roots of
$f'$.
\item\label{item:outder} Output (in parallel) each $\alpha_j$.
\item Put $c=2+\max_{j<d}\lh{f_j/f_d}$ and
$k_{\max}=\cl{\log(2c\ep^{-1})/\log A}$.
\item\label{item:loop} For every $j<s$, $k<k_{\max}$, and $q<p$,
do the following in parallel.
\item\label{item:parms} Let $a=\alpha_j+\ep A^k\xi^q$, $b=f(a)$, $R=\frac12\lh{a-\alpha_j}$,
$N=\cl{\log_2(\mu R\ep^{-1})}$.
\item For each $h\le d$, let $\tilde f_h=\sum_{u=h}^d\binom
uhf_ua^{u-h}$.
\item\label{item:last} Compute and output
$$z_{j,k,q}=a+\sum_{\substack{m_2,\dots,m_d\\\sum_h(h-1)m_h<N}}
 \frac{(2m_2+\dots+dm_d)!\,(-\tilde f_2)^{m_2}\cdots(-\tilde f_d)^{m_d}
       (-b)^{1+m_2+\dots+(d-1)m_d}}
  {m_2!\cdots m_d!\,(1+m_2+\dots+(d-1)m_d)!\,\tilde f_1^{1+2m_2+\dots+dm_d}}.$$
\end{enumerate}

Let $f(\alpha)=0$, we have to show that one of the numbers output by
the algorithm is $\ep$-close to $\alpha$. If
$\lh{\alpha-\alpha_j}<\ep$ for some $j$, we are
done by step \eqref{item:outder}. We can thus assume
$\dist(\alpha,C)\ge\ep$, which implies $\dist(\alpha,C_f)\ge3\ep/4$.
Assume that $\alpha_j$ is an $\ep/4$-approximation of the root $\tilde\alpha_j$
of $f'$
nearest to $\alpha$. Since all roots of $f$ or $f'$ have modulus
bounded by $c-1$ by Cauchy's bound, we have
$\ep\le\lh{\alpha-\alpha_j}<2c$, thus there exists $k<k_{\max}$ such
that $\ep A^k\le\lh{\alpha-\alpha_j}<\ep A^{k+1}$. Let
$q<p$ be such that the argument of $\alpha-\alpha_j$ differs from
$2\pi q/p$ by at most $\pi/p$, and consider steps
\eqref{item:loop}--\eqref{item:last} for this particular choice of
$j,k,q$ (cf.\ Fig.~\ref{fig:pts}).
\begin{figure}
\begin{center}
\includegraphics*{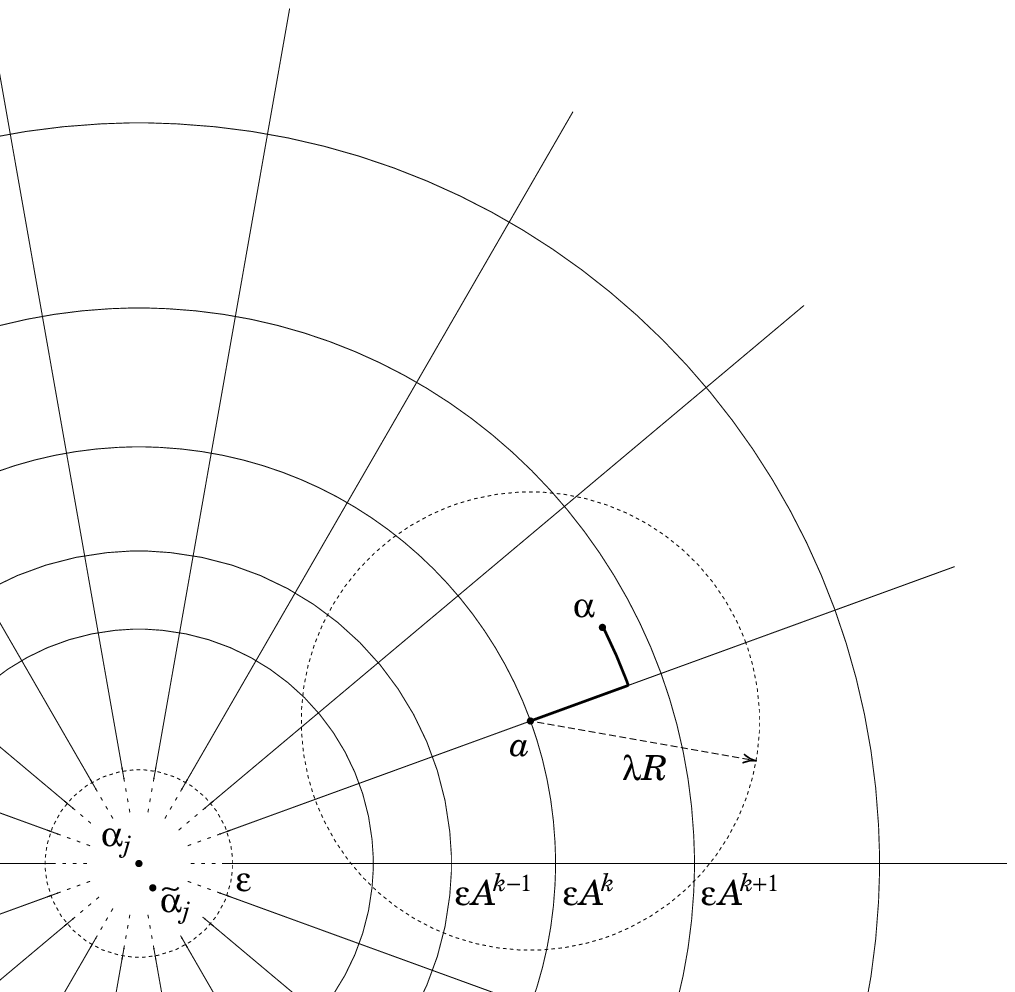}
\end{center}
\caption{The spiderweb.}
\label{fig:pts}
\end{figure}
We have
$$\lh{\alpha-a}\le\left(\frac\pi p+1-\frac1A\right)\lh{\alpha-\alpha_j}
\le\frac{2\lambda}5\,\lh{\alpha-\alpha_j}<\frac15\,\lh{\alpha-\alpha_j}.$$
Notice that
$$\dist(\alpha,C_f)=\lh{\alpha-\tilde\alpha_j}\ge\lh{\alpha-\alpha_j}-\frac\ep4$$
by the choice of $\tilde\alpha_j$ and $\alpha_j$, hence
$$
\dist(a,C_f)\ge\dist(\alpha,C_f)-\lh{a-\alpha}
\ge\lh{\alpha-\alpha_j}-\frac\ep4-\frac15\,\lh{\alpha-\alpha_j}
\ge\frac12\,\lh{\alpha-\alpha_j}\ge R.
$$
Since
$$\lh{a-\alpha_j}\ge\lh{\alpha-\alpha_j}-\lh{a-\alpha}
>\frac45\,\lh{\alpha-\alpha_j},$$
we also have
$$\lh{a-\alpha}<\frac54\,\frac{2\lambda}5\,\lh{a-\alpha_j}=\lambda R.$$
Let
$$g(w)=a+\sum_{n=1}^\infty c_n(w-b)^n$$
be an inverse of $f$ in a neighbourhood of $b$, and let $\roo$ be its
radius of convergence. By \th\ref{thm:radius},
$\lh{-b}=\lh{f(\alpha)-b}<\roo_0/2$, where $\roo_0=\nu R\,\lh{f'(a)}\le\roo$.
Thus, $g(f(\alpha))=g(0)=\alpha$. Since $\sum_h\tilde f_hz^h=f(z+a)$ by the
binomial formula, $\tilde f_h=[(z-a)^h]f$. Then it follows from
\th\ref{thm:invser} that
$$z_{j,k,q}=a+\sum_{n=1}^Nc_n(-b)^n.$$
Since
$$\lh{c_n(-b)^n}\le\frac{\mu R}{n\roo_0^n}\lh b^n
<\frac{\mu R}{2^n}$$
by \th\ref{thm:radius}, we have
$$\lh{\alpha-z_{j,k,q}}=
\biggl|\sum_{n=N+1}^\infty c_n(-b)^n\biggr|
<\frac{\mu R}{2^N}\le\ep.\qedhere$$
\end{Pf}
Most of the algorithm described in \th\ref{thm:list} is independent of
the assumption of $d$ being constant (or it can be worked around).
There are two principal exceptions. First, the recursion in step
\eqref{item:rec} amounts to $d$ sequential invocations of the
algorithm. Second, while $N$ is still linear in the size of the input,
the main sum in step \eqref{item:last} has roughly $N^d$ terms. Thus,
approximation of roots of arbitrary univariate
polynomials can be done by (uniform) threshold circuits of depth
$O(d)$ and size $n^{O(d)}$, where $n$ is the total length of the input.
(The known $\cxt{NC}$ algorithms for root finding can do much better for
large $d$.)

The algorithm from \th\ref{thm:list} does the hard work in locating the
roots of $f$, but it suffers from several drawbacks:
\begin{itemize}
\item Its output includes a lot of bogus results that are not
actually close to any root of $f$.
\item There may be many elements on the list close to the same root,
and we do not get any information on the multiplicity of the roots.
\item The roots have no ``identity'': if we run the algorithm for two
different $t$s, we do not know which approximate roots on the output lists
correspond to each other.
\item It may be desirable to output the binary expansions of the roots
rather than just approximations.
\end{itemize}
We are going to polish the output of the algorithm to fix these problems. Let us first
formulate precisely the goal.
\begin{Def}\th\label{def:goal}
The \emph{$t$-digit binary expansion} of $a\in\C$ is the pair
$\p{\fl{\Re(a2^t)},\fl{\Im(a2^t)}}$, where both integers are written
in binary. A \emph{root-finding algorithm} for a set of polynomials
$P\sset(\Q(i))[x]$ is an algorithm with the following properties:
\begin{enumerate}
\item The input consists of a polynomial $f\in P$ given by a
list of its coefficients in binary, and a positive integer $t$ in unary.
\item The output is a list of pairs
$\{\p{z_j(f,t),e_j(f,t)}:j<s(f,t)\}$.
\item For every $f\in P$, there exists a factorization
$$f(z)=c\prod_{j<s}(z-a_j)^{e_j},$$
where $c\in\Q(i)$, $a_j\in\C$, $a_j\ne a_k$ for $j\ne k$, and $e_j>0$,
such that for every $t$: $s(f,t)=s$, $e_j(f,t)=e_j$, and $z_j(f,t)$ is
the $t$-digit binary expansion of $a_j$.
\end{enumerate}
We note that the choice of base $2$ in the output is arbitrary, the
algorithm can output expansions in any other base if needed.
\end{Def}

\begin{Lem}\th\label{lem:sq-free}
Let $d$ be a constant. Given a degree $d$ polynomial $f\in(\Q(i))[x]$,
we can compute in uniform $\tc$ a list of pairwise coprime square-free
nonconstant polynomials $f_j$, $c\in\Q(i)$, and integers $e_j>0$ such that
$f=c\prod_{j<k}f_j^{e_j}$, where $k,e_j\le d$.
\end{Lem}
\begin{Pf}
Since $d$ is constant, division of degree $d$ polynomials takes $O(1)$
arithmetical operations, hence it can be implemented in uniform $\tc$.
The same holds for $\gcd$, using the Euclidean algorithm.
We compute a list
$L=\p{f_j:j<k}$, $k\le d$, of nonconstant polynomials such that
$f=\prod_jf_j$ as follows:
\begin{enumerate}
\item Start with $L=\p{f}$. Repeat the following steps until none of
them is applicable.
\item If $f_j$ is not square-free, replace it with $\gcd(f_j,f'_j)$ and
$f_j/\gcd(f_j,f'_j)$.
\item If $f_h\divi f_j$, $f_j\nmid f_h$ for some $h,j$, replace $f_j$ in
$L$ with $f_h,f_j/f_h$.
\item If $g:=\gcd(f_h,f_j)\ne1$ for some $h,j$ such that $f_h\nmid
f_j$, $f_j\nmid f_h$,
replace $f_h,f_j$ in $L$ with $g,g,f_h/g,f_j/g$.
\end{enumerate}
The algorithm terminates after at most $d$ steps, hence it is
in~$\tc$. Clearly, it computes a list of square-free polynomials such
that for every $h,j$, $f_h$ is coprime to $f_j$ or $f_h$ is a scalar
multiple of $f_j$. It remains to collect scalar multiples of the
same polynomial together.
\end{Pf}
\begin{Lem}\th\label{lem:weed}
Let $d$ be a constant. Given a degree $d$ square-free polynomial
$f\in(\Q(i))[x]$ and $t$ in unary, we can compute in uniform $\tc$ a list
$\{z_j:j<s\}$ such that every root of $f$ is within distance $2^{-t}$
of some $z_j$, and every $z_j$ is within distance $2^{-t}$ of some root.
\end{Lem}
\begin{Pf}
We use the notation from the proof of \th\ref{thm:list}.
We modify the algorithm from that proof as
follows:
\begin{itemize}
\item We compute an $\ep_0>0$ such that the distance of any root of
$f$ to any root of $f'$ is at least $\ep_0$ using \th\ref{lem:sep}. In
step \eqref{item:rec}, we put $\ep=\min(2^{-t},\ep_0/3)$.
\item We skip step \eqref{item:outder}.
\item In step \eqref{item:parms}, we check that $\lh
b<\frac12\nu\lh{f'(a)}R$ and $\lh{a-\alpha_{j'}}\ge R+\ep/4$ for every
$j'<s$. If either condition is violated, we output a symbol ``$*$''
instead of a number, and skip the remaining two steps.
\end{itemize}
The result is a list of numbers and $*$'s; it is easy to construct the
sublist consisting of only numbers by a $\tc$ function.

Let $z_{j,k,q}$ be one of the numbers output by the algorithm. In step
\eqref{item:parms} we ensured $\dist(a,C)\ge R+\ep/4$, hence
$\dist(a,C_f)\ge R$. Moreover, $\lh{0-b}<\roo_0/2$, hence $0$ is
within the radius of convergence of $g$, and $\alpha=g(0)$ is a root
of $f$ whose distance from $z_{j,k,q}$ is
$$\lh{\alpha-z_{j,k,q}}=
\biggl|\sum_{n=N+1}^\infty c_n(-b)^n\biggr|
<\frac{\mu R}{2^N}\le\ep.$$

On the other hand, let $\alpha$ be a root of $f$. Since
$\dist(\alpha,C_f)\ge\ep_0$, we have $\dist(\alpha,C)\ge\ep$, hence we
can choose $j,k,q$ such that $\lh{\alpha-z_{j,k,q}}<\ep$ as in the
proof of \th\ref{thm:list}. We have to show that the extra conditions
in step \eqref{item:parms} are satisfied. $\lh b<\frac12\nu
R\lh{f'(a)}$ was verified in the proof of \th\ref{thm:list}. Moreover,
$$\lh{a-\alpha_{j'}}\ge
  \lh{\alpha-\tilde\alpha_{j'}}-\lh{a-\alpha}-\frac\ep4
 \ge\lh{a-\tilde\alpha_j}-\lh{a-\alpha}-\frac\ep4
 \ge\frac45\lh{\alpha-\alpha_j}-\frac\ep2\ge R+\frac\ep4$$
as $\lh{\alpha-\alpha_j}\ge\ep_0-\ep/4>\frac52\ep$.
\end{Pf}
We can now finish the proof of the main result of this paper:
\begin{Thm}\th\label{thm:root}
For every constant $d$, there exists a uniform $\tc$ root-finding
algorithm for degree $d$ polynomials in the sense of \th\ref{def:goal}.
\end{Thm}
\begin{Pf}
We employ the notation of \th\ref{def:goal}.
By \th\ref{lem:sq-free}, we can assume $f$ to be square-free
(in which case we will have $e_j(f,t)=1$ for all $j$, so we only need to
compute the roots). Consider the following $\tc$ algorithm:
\begin{enumerate}
\item Using \th\ref{lem:sep}, compute an $\eta>0$ such that all roots
of $f$ are at distance at least $\eta$ from each other.
\item Using \th\ref{lem:weed}, compute a list $\{r'_j:j<u\}$ such that
every root of $f$ is within distance $\eta/5$ of some $r'_j$, and vice
versa.
\item Note that if $r'_h$ and $r'_j$ correspond to the same root, then
$\lh{r'_h-r'_j}<\frac25\eta$, otherwise $\lh{r'_h-r'_j}>\frac35\eta$.
Use this criterion to omit duplicate roots from the list, creating a list
$\{r_j:j<d\}$ which contains $\eta/5$-approximations of all roots of
$f$, each of them exactly once.
\item If $\ep:=2^{-t}\ge\eta/5$, output $z_j:=r_j$ and halt. Otherwise use
\th\ref{lem:weed} to construct a list $\{z'_h:h<s\}$ consisting of
$\ep$-approximations of roots of $f$.
\item For each $j<d$, output $z_j:=z'_{h(j)}$, where $h(j)$ is the smallest
$h<s$ such that $\lh{z'_h-r_j}<\eta/2$.
\end{enumerate}
Notice that the computation of $r_j$ is independent of $t$. Let $a_j$ be the unique root of $f$ such that
$\lh{a_j-r_j}<\eta/5$. Given $t$ and $i$, let $j'$ be such that
$\lh{z'_h-a_{j'}}<\ep$. Then
$\lh{z'_h-r_j}<\ep+\eta/5\le\frac25\eta$ if $j=j'$,
otherwise $\lh{z'_h-r_j}>\frac45\eta-\ep\ge\frac35\eta$. Thus,
the definition of $h(j)$ in the last step is sound, and guarantees
$\lh{z_j-a_j}<\ep$.

It follows that this $\tc$ function has all the required properties,
except that it computes approximations instead of binary expansions.
We can fix this as follows. Using the algorithm we have just described, we can compute
integers $u,v$ such that $\lh{u+iv-2^ta_j}<1$. Then $\fl{\Re(2^ta_j)}$
is either $u$ or $u-1$, hence it remains to find the sign of
$\Re(2^ta_j)-u$ (the case of $\Im$ is similar).

Let $g(z)=f(2^{-t}(2z+u))$, $h(z)=\ob g(-z)$, and
$\alpha=\frac12(2^ta_j-u)$. Then $g(\alpha)=0=h(-\ob\alpha)$ and
$\alpha-(-\ob\alpha)=\Re(2^ta_j)-u$. Using \th\ref{lem:sep}, we can
compute $\xi>0$ such that $\lh{\alpha-(-\ob\alpha)}\ge\xi$ whenever
it is nonzero. Using the algorithm above,
we can compute rational $u',v'$ such that $\lh{u'+iv'-2^ta_j}<\xi/4$.
If $\lh{u-u'}<\xi/2$, then $\Re(2^ta_j)=u$. Otherwise,
$\lh{\Re(2^ta_j)-u}\ge\xi$, hence the sign of $u'-u$ agrees with the
sign of $\Re(2^ta_j)-u$.
\end{Pf}
\begin{Cor}\th\label{cor:algnum}
If $\alpha$ is a fixed real algebraic number, then the $k$th
bit of $\alpha$ can be computed in uniform $\tc$,
given $k$ in unary.
\noproof\end{Cor}
(Note that this corollary is only interesting in the uniform setting,
since the language is unary.)

\section{Open induction in $\vtc$}\label{sec:open-induction}
As we already mentioned in the introduction, our primary motivation
for studying root finding for constant-degree polynomials comes from
bounded arithmetic. We will now describe the connection in more
detail. A reader not interested in bounded arithmetic may safely stop
reading here.

The basic objects of study in bounded arithmetic are weak first-order
theories based on integer arithmetic. There is a loose correspondence
of arithmetical theories to complexity classes: in particular, if a
theory $T$ corresponds to a class $C$, then the provable total
computable functions of $T$ are functions from $C$ (or more precisely,
$FC$). The following is one of the natural problems to study in this
context: assume we have a concept (say, a language or a function) from
the computational class $C$. Which properties of this concept are
provable in the theory $T$? (This asks for a form of feasible
reasoning: what can we show about the concept when we are restricted
to tools not exceeding its complexity?)

Here we are concerned with the theory $\vtc$, corresponding to $\tc$.
We refer the reader to Cook and Nguyen \cite{cook-ngu} for a
comprehensive treatment of $\vtc$. Let us briefly recall that $\vtc$
is a two-sorted theory, with one sort intended for natural numbers
(which we think of as given in unary), and one sort for finite sets of
these unary numbers (which we also regard as finite binary strings, or
as numbers written in binary). We are primarily interested in the
binary number sort, we consider the unary sort to be auxiliary. We use
capital letters $X,Y,\dots$ for variables of the binary (set) sort,
and lowercase letters $x,y,\dots$ for the unary sort. The language of
the theory consists of basic arithmetical operations on the unary
sort, the elementhood (or bit) predicate $x\in X$, and a function $\lh
X$ which extracts an upper bound on elements of a set $X$. The axioms
of $\vtc$ include comprehension for $\Sig0$ formulas (formulas with
number quantifiers bounded by a term and no set quantifiers)---which
also implies induction on unary numbers for $\Sig0$ formulas---and an
axiom ensuring the existence of counting functions for any set. The
provably total computable (i.e., $\Sigma^1_1$-definable: $\Sigma^1_1$
formulas consist of a block of existential set quantifiers in front of
a $\Sig0$ formula) functions of $\vtc$ are the $\tc$ functions.

In $\vtc$, we can define the basic arithmetical operations
$+,\cdot,\le$ on binary integers. Our main question is, what
properties of these operations are provable in $\vtc$. (We can make
this more precise as follows: which theories in the usual
single-sorted language of arithmetic $L_\PA=\p{0,1,+,\cdot,\le}$ are
interpreted in $\vtc$ by the corresponding operations on the binary
sort?) It is not hard to show that $\vtc$ proves binary integers to
form a discretely ordered ring ($\M{DOR}$). What we would especially like to know
is whether $\vtc$ can prove the induction schema on the binary sort
$$\fii(0)\land\forall X\,(\fii(X)\to\fii(X+1))\to\forall X\,\fii(X)$$
for some nontrivial class of formulas $\fii$. In particular, we want to know whether
$\vtc$ includes the theory $\io$ (axiomatized by induction for open
formulas of $L_\PA$ over $\M{DOR}$) introduced by Shepherdson
\cite{sheph} and widely studied in the literature.

Now, assume for a moment that $\vtc\vdash\io$. Then for each constant
$d$, $\vtc$ proves
$$X<Y\land F(X)\le0<F(Y)\to\exists Z\,(X\le Z<Y\land F(Z)\le0<F(Z+1))$$
where $F(X)=\sum_{j\le d}U_jX^j$ is a degree $d$ integer
polynomial whose coefficients are parameters of the formula. This is
(equivalent to) a $\Sigma^1_1$ formula, hence the existential
quantifier is, provably in $\vtc$, witnessed by a $\tc$ function
$G(U_0,\dots,U_d,X,Y)$. Since any rational polynomial is a scalar
multiple of an integer polynomial, and we can pass from a polynomial
$F(X)$ to $2^{td}F(2^{-t}X)$ to reduce the error from $1$ to $2^{-t}$,
we see that there is a $\tc$ algorithm solving the following
root-finding problem: given a degree $d$ rational polynomial and two
rational bounds where it assumes opposite signs, approximate a real root of
the polynomial between the two bounds up to a given accuracy. Using a
slightly more complicated argument, one can also obtain a root-finding
algorithm in the set-up we considered earlier : i.e., we approximate all
complex roots of the polynomial, and the input of the algorithm is
only the polynomial and the desired error of approximation. Thus, a
$\tc$ root-finding algorithm is a necessary prerequisite for showing
$\io$ in $\vtc$.

We can in a sense reverse the argument above to obtain a proof of open
induction from a root-finding algorithm, but there is an important
caveat. The way we used the witnessing theorem for $\vtc$, we lost the
information that the soundness of the algorithm is provable in $\vtc$.
Indeed, if we are only concerned with the computational complexity of
witnessing functions, then witnessing of $\Sigma^1_1$ formulas is
unaffected by addition of true universal (i.e., $\forall\Sig0$) axioms
to the theory. In other words, the same argument shows the existence
of a root-finding algorithm from the weaker assumption
$\vtc+\Th_{\forall\Sigma^B_0}(\stm)\vdash\io$, where
$\Th_{\forall\Sigma^B_0}(\stm)$ denotes the set of all $\forall\Sigma^B_0$
sentences true in the standard model of arithmetic. Now, this
formulation of the argument can be reversed:

\begin{Thm}\th\label{thm:vtc+pi1}
The theory $\vtc+\Th_{\forall\Sigma^B_0}(\stm)$ proves $\io$ for the
binary number sort.
\end{Thm}
\begin{Pf}
Let $M$ be a model of $\vtc+\Th_{\forall\Sigma^B_0}(\stm)$, and $D$
be the discretely ordered ring of the binary integers of $M$.
For any constant $d$, we can use \th\ref{thm:list} to
construct a $\tc$ function which, given
the coefficients of an integer polynomial of degree $d$, computes a
list of integers $a_0<a_1<\dots<a_k$, $k\le d$, such that the sign of
the polynomial is constant on each of the \emph{integer} intervals
$(a_j,a_{j+1})$, $(-\infty,a_0)$, $(a_k,+\infty)$. This property of
the function is expressible by a $\forall\Sig0$ sentence (when the
coefficients of the polynomial and the $a_j$ are taken from the binary
sort), hence it holds in $D$ that such elements $a_0,\dots,a_k$ exist
for every polynomial over $D$.

Any atomic formula $\fii(x)$ of $L_\PA$ with parameters from $D$ is
equivalent in $\M{DOR}$ to the formula $f(x)\le0$ for some $f\in
D[x]$, hence $\fii(D):=\{x:D\models\fii(x)\}$ is a finite union of
intervals. Sets of this kind form a Boolean algebra, hence
$\fii(D)$ is a finite union of intervals for every open formula
$\fii$. This implies induction for $\fii$: if
$D\models\fii(0)\land\neg\fii(u)$ for some $u>0$, the interval $I$ of
$\fii(D)$ containing $0$ cannot be infinite from above, hence its
larger end-point $v\in D$ satisfies
$D\models\fii(v)\land\neg\fii(v+1)$.
\end{Pf}
\begin{Prob}\th\label{prob:vtc-iopen}
Does $\vtc$ prove $\io$?
\end{Prob}
In light of the discussion above, \th~\ref{prob:vtc-iopen} is essentially
equivalent to the following: are there $\tc$ root-finding
algorithms for constant-degree polynomials \emph{whose correctness is
provable in $\vtc$}? We remark that the complex-analytic tools we used in
the proof of \th\ref{thm:list} are not available in $\vtc$.

We note that already proving the totality of integer division in
$\vtc$ (i.e., formalization of a $\tc$ integer division
algorithm in $\vtc$) is a nontrivial open\footnote{Hesse et
al.~\cite[Cor.~6.6]{hab} claim that the totality of integer division is
provable in $\vtc$ (or rather, in the theory $C^0_2$ of Johannsen and
Pollett \cite{joh-pol:c02}, RSUV-isomorphic to
$\vtc+\Sig0$-\ac, which is $\forall\Sigma^1_1$-conservative over $\vtc$). However,
the way it is stated there with no proof as an ``immediate'' corollary
strongly suggests that the claim is due to a misunderstanding.
See also \cite[\S IX.7.3]{cook-ngu}.} problem, thus
\th\ref{prob:vtc-iopen} may turn out to be too ambitious
a goal. The following is a still interesting version of the question,
which may be easier to settle:
\begin{Prob}\th\label{prob:vtc-imul}
Does $\vtc+\M{IMUL}$ prove $\io$, where $\M{IMUL}$ is a natural axiom
postulating the totality of iterated integer multiplication?
\end{Prob}

We also mention that it is not hard to prove in~$\vtc$ that binary
integers form a $\Z$-ring, which implies all universal
consequences of~$\io$ in the language of ordered rings. The problem
is thus only with statements with a genuinely existential import (note
that $\io$ is a $\forall\exists$ theory).

\subsection*{Acknowledgements}
I am grateful to Paul Beame and Yuval Filmus for useful discussions,
and to anonymous referees for helpful suggestions.

\bibliographystyle{mybib}
\bibliography{mybib}
\end{document}

%% file: defs.tex
\ifx\mydefs\undefined\else \fi
\let\mydefs\relax

\allowdisplaybreaks[2]

\ifx\loadcyr\undefined\else
\input cyracc.def
\makeatletter
 at 1\@ptsize pt
 at 1\@ptsize pt
 at 1\@ptsize pt
\makeatother

\fi

\let\M\mathit
\def\gobble#1{}
\def\fixsup#1#2{{#1\let\dp\gobble\mathstrut}^#2_}

\def\bme{\hskip.75em\relax}

\def\?{\mathbin?}

\newbox\circlebox
\setbox\circlebox\hbox{$\bigcirc$}
\def\circled#1{%
  \setbox0\hbox to\wd\circlebox{\hss$#1$\hss}\wd0=0pt
  \box0\copy\circlebox}

\let\fii\varphi
\let\tet\vartheta
\let\ep\varepsilon

\let\roo\varrho
\def\greek#1{$\expandafter\greeknum\csname c@#1\endcsname$}
\makeatletter
\def\greeknum#1{\ifcase#1\or\alpha\or\beta\or\gamma\or\delta\or\ep
      \or\digamma\or\zeta\or\eta\or\tet\or\iota\else\@ctrerr\fi}
\makeatother

\def\p#1{\langle#1\rangle}

\def\lh#1{\lvert#1\rvert}

\let\bez\smallsetminus

\let\sset\subseteq

\let\Sset\supseteq

\let\onto\twoheadrightarrow

\let\nul\varnothing
\def\res{\mathbin\restriction}

\def\twoprimes{\raise.2\fontdimen6\scriptfont2\hbox{$\scriptstyle\prime\prime$}}
\newcommand\rpair[3][3em]{\mathrel{%
   \begin{matrix}%
     \strut\smash{\xrightonto{\hbox to#1{\hss$#2$\hss}}}\\[-1.7ex]%
     \strut\smash{\xleftembed[\hbox to#1{\hss$#3$\hss}]{}}%
   \end{matrix}}}
\makeatletter
\newcommand\xrightonto[2][]{\ext@arrow 0359\rightontofill{#1}{#2}}
\newcommand\xleftembed[2][]{\ext@arrow 3095\leftembedfill{#1}{#2}}
\def\leftembedfill{\arrowfill@\leftarrow\relbar\hookleftnoarrow}
\def\rightontofill{\arrowfill@\relbar\relbar\onto}
\def\hookleftnoarrow{\DOTSB\relbar\joinrel\rhook}
\makeatother

\def\fl#1{\lfloor#1\rfloor}

\def\cl#1{\lceil#1\rceil}

\mathchardef\#="2023 

\let\divi\mid

\ifx\busspf\undefined\else
\usepackage{bussproofs}
\EnableBpAbbreviations

\fi

\ifx\symlasy\undefined

  \def\centerdot#1{%
    \setbox0\hbox{$\mathop{#1}$}\dimen0 \ht0
    \setbox0\hbox{$#1$}\advance\dimen0 -\ht0
    \setbox2\hbox to\wd0{\hss$\mathop{\cdot}$\hss}\wd2=0pt
    \lower\dimen0\box2\box0 }
\else
  
  \def\centerdot#1{%
     \setbox0\hbox{$#1$}%
     \raise0.206\ht0\hbox to\wd0{\hss$\cdot$\hss}%
     \kern-\wd0 \box0 }
\fi

\let\sls|

\def\Up{{\setbox0\hbox{$\uparrow$}%
         \lower\dp0\hbox to\wd0{\hss\vrule width4pt height.4pt\hss}%
         \kern-\wd0\box0}}
\def\UP{{\setbox0\hbox{$\uparrow$}%
         \lower\dp0\hbox to\wd0{\hss\vrule width4pt height.4pt\hss}%
         \kern-\wd0\copy0\kern-\wd0\raise.35ex\box0}}
\def\Down{{\setbox0\hbox{$\downarrow$}%
         \raise\ht0\hbox to\wd0{\hss\vrule width4pt depth.4pt\hss}%
         \kern-\wd0\box0}}

\DeclareMathOperator\bit{bit}

\DeclareMathOperator\id{id}

\DeclareMathOperator\Th{Th}

\DeclareMathOperator\Int{int}

\DeclareMathOperator\Res{Res}

\DeclareMathOperator\dist{dist}
\let\Re\relax
\let\Im\relax
\DeclareMathOperator\Re{Re}
\DeclareMathOperator\Im{Im}

\let\cxt\mathrm

\def\ptime{\cxt P}

\def\nc{\cxt{NC}}

\def\tc{\cxt{TC}^0}
\def\Ac{\cxt{AC}^0}

\def\Sig{\Sigma^B_}

\def\st{\expandafter\hat}
\def\io{\M{IOpen}}

\def\vtc{\M{VTC}^0}

\def\ac{$\M{AC}$}

\def\PA{\M{PA}}

\def\Q{\mathbb Q}
\def\Z{\mathbb Z}
\def\RR{\mathbb R}
\def\C{\mathbb C}

\let\stm\N

\mathcode`\*="0203
\let\ob\overline

\mathchardef\mhyphen="2D

\def\noproof{\leavevmode\unskip\bme\vadjust{}\nobreak\hfill$\qed$\par}
\let\qed\Box
\newenvironment{Pf}[1][]
  {\par\noindent\textit{Proof\optpar{#1}:}\bme\ignorespaces}
  {\noproof\pagebreak[2]\vskip\medskipamount\ignorespacesafterend}
\def\qedhere{\relax\ifmmode\eqno\qed\expandafter\aftergroup
                   \else\noproof\fi\noqed}
\def\noqed{\let\noproof\relax}

\theoremstyle{plain}
\ifx\shortthm\undefined
\newtheorem{Thm}{Theorem}[section]
\else
\newtheorem{Thm}{Theorem}
\fi
\newtheorem{Prop}[Thm]{Proposition}
\newtheorem{Cor}[Thm]{Corollary}
\newtheorem{Lem}[Thm]{Lemma}
\newtheorem{Fact}[Thm]{Fact}

\newtheorem{Prob}[Thm]{Problem}

\def\theCl{\arabic{Cl}}

\theorembodyfont\upshape
\newtheorem{Def}[Thm]{Definition}
\newtheorem{Rem}[Thm]{Remark}
\newtheorem{Exm}[Thm]{Example}

\newenvironment{Pf*}{\let\qed\qedCl\Pf}\endPf

\usepackage[reftex]{theoremref}
\newif\iflinenumbers
\linenumberstrue

{\catcode`\^^I=13 \catcode`\^^M=13
\gdef\doalgo#1#2\end#{\hbox to\hsize{\hss \let^^I\qquad%
  \def\\^^M{\nobreak\hfil\break\vadjust{}\qquad}%
  \fboxsep1em \linenum0 %
  \fbox{\hsize#1\vbox{%
  \everypar{\advance\linenum1 %
      \hbox to1.2em{%
           \hss\iflinenumbers$\scriptstyle\the\linenum$\hskip.6em\fi}}%
  #2}}\hss}\end}}
\newcount\linenum

\def\key{\relax\ifmmode\expandafter\mathbf\else\expandafter\textbf\fi}

\def\allowhyphens{\nobreak\hskip0pt\relax}

\DeclareRobustCommand*\magiclparen{\ifmmode(\else\textup(\allowhyphens\fi}
\DeclareRobustCommand*\magicrparen{\ifmmode)\else\textup)\fi}
\let\lparen=(  \let\rparen=)
\def\magicparon{\catcode`\(\active\catcode`\)\active}
\def\magicparoff{\catcode`\(12 \catcode`\)12 }
\AtBeginDocument{\ifx\ifPreview\iftrue\else\magicparon\fi}
\magicparon
\let (=\magiclparen  \let )=\magicrparen

\ifx\enumup\undefined
  
\else
  
\fi

\def\optpar#1{\ifx\relax#1\relax\else\/ (#1)\fi}

\magicparoff

\def\cmb{\mathcode`\,"8000 }
\mathchardef\comma=\mathcode`\,
{\catcode`\,=\active \gdef,{\comma\penalty\relpenalty}}

\providecommand\dedic{\thanks{Supported by
grant IAA100190902 of GA AV \v CR, Center of Excellence CE-ITI under the grant
P202/12/G061 of GA \v CR, and RVO: 67985840.}}
\author{Emil Je\v r\'abek\dedic\\[\medskipamount]
Institute of Mathematics of the Academy of Sciences\\
\small \v Zitn\'a 25,
115\:67 Praha 1,
Czech Republic,
email: \texttt{jerabek@math.cas.cz}
}